\documentstyle[preprint,prb,aps]{revtex}
 \title{Negative linear compressibility in confined dilatating systems 
}

 \author{E.V. Vakarin$^a$, Yurko Duda$^b$, J.~P.~Badiali$^a$
}
 \address{
$^a$ UMR 7575 LECA ENSCP-UPMC, 11 rue P. et M. Curie, 75231 Cedex 05, Paris, 
France\\ 
$^b$ Programa de Ingenier\'ia Molecular, Instituto Mexicano del Petr\'oleo,
07730 D. F., M\'exico}

\begin{document}
 \maketitle
 \begin{abstract}
The role of a matrix response to a fluid insertion is analyzed
in terms of a perturbation theory and Monte Carlo simulations applied to a 
hard sphere fluid in a slit of fluctuating density-dependent width. 
 It is demonstrated that a coupling of the fluid-slit repulsion, spatial 
 confinement and the matrix dilatation acts as an effective 
 fluid-fluid attraction, inducing a pseudo-critical 
state with divergent linear compressibility and non-critical density 
fluctuations. An appropriate combination of the dilatation rate, 
fluid density and the slit size leads to the fluid states with 
negative linear compressibility. It is shown that the switching from 
positive to negative compressibility is accompanied by  an abrupt change in 
the packing mechanism. 

\end{abstract}

%\begin{multicols}{2}

\section{Introduction}
Compounds with a negative linear (or surface) compressibility 
have recently attracted an interest\cite{Baughman1,Baughman2,Scicomment} 
because of their specific properties (stretch-induced densification and 
auxetic behavior), which might have promising applications. In the case of 
pure materials only "rare" crystal phases exhibit these effects, while for 
composite structures\cite{Bowick} it seems to be rather common. Recent 
experimental studies on insertion into organic\cite{EXP} and 
non-organic\cite{Jornada} matrices reveal the negative compressibility 
effects due to the host-guest coupling. Conceptually similar escape 
transition\cite{escape} occurs when a polymer chain is compressed between 
two pistons. Charge separation in confined fluids has also been
accompanied\cite{Lozada} with the negative compressibility.  
These examples suggest that the insertion 
systems with confinement and swelling (dilatation) should  exhibit this 
generic effect at appropriate conditions. In particular, two-particle 
systems confined to two-dimensional finite-size boxes of different geometry 
have been studied\cite{Two1,Two2,Two3} in this context. It has been found 
that the isotherms exhibit a van der Waals-type (vdW) instability (loop) as 
the box size passes through a critical value. The instability has been 
associated with a prototype of a liquid-gas or a liquid-solid transition in 
many-body systems. A quite similar instability appears in two-dimensional 
granular media\cite{Argentina}, whose effective temperature decreases with 
the density, leading to a phase separation.  Similar features have been 
detected\cite{Mukamel} in hard discs confined to a narrow channel. 
Nevertheless, the non-monotonic behavior has been attributed to a sharp 
change in the accessible phase space, without any connection to collective 
effects typical for the conventional phase transitions.  It is well-known, 
however, that a liquid-gas transition in finite systems\cite{Gulminelli} 
manifests itself through the negative compressibility states due to the 
surface effects.

Thus, it is quite interesting to find out  if such a loop could exist 
in three-dimensional many-body systems and what is the physics behind.
In particular, we focus on a prototype of an insertion system.
One of the key features of such systems is a host dilatation (or 
contraction) upon a guest accommodation. This is evident from experiments 
with high-porosity materials, like aerogels\cite{aerogel1,aerogel}. Upon 
adsorption such matrices change in volume and their pore size distribution 
depends on the adsorbate pressure (or density). This effect is not exclusive 
to relatively soft gel-like matrices, it is quite common for  carbon 
nanotubes\cite{Cole}  and various intercalation 
compounds\cite{PRB,JPCB,SOSI}. In anisotropic cases (e.g. layered 
compounds) one deals with a competition of two effects. An increase of the 
lateral dimension (stretching) at constant number of particles tends to 
decrease the guest density. This usually induces a transversal shrink, 
leading to a densification in this direction. Since we have a composite 
host-guest system, the shrink does not obey the linear elasticity rules and, 
depending on the shrink intensity, one might expect the negative 
compressibility. In particular, it has been shown\cite{SOSI} that a 
nonlinear increase of the host size with the guest concentration may induce 
a liquid-gas coexistence for the guest fluid even if the bulk liquid phase 
does not exist (e.g. a hard-sphere fluid). This means that the negative 
compressibility states were indeed present, but they were unstable under the 
conditions imposed.

Therefore, the objective of this paper is to find the conditions at which
a coupling of the host dilatation  and the guest confinement
can stabilize the negative compressibility states.  
For this purpose we start with a quite simple model - the hard spheres,
adsorbed in a softly-repulsing planar slit with a density-dependent width. 
The model contains all the relevant features: confinement, 
dilatation and spatial anisotropy.  The latter is important because 
it offers a possibility of observing negative linear and positive tangential 
compressibilities, preserving the overall thermodynamic stability. 
On the other hand, in the absence of the dilatation  (fixed 
slit width), the model is phenomenologically simple\cite{Wu,Ale}. In 
particular, there is no a liquid-vapor transition (both in the bulk and in 
confined geometries). 
Such that the new 
aspects are not masked by the internal complexity. 
The system is analyzed by 
means
of Monte Carlo (MC) simulation and a first-order perturbation 
theory\cite{Schoen,Kuz,pore}.
Qualitative reliability of this approach 
has been tested in
application to the liquid-vapor coexistence\cite{Kuz} 
and phase
separation\cite{pore} in confined geometries.

%%%%%%%%%%%%%%%%%%%%%%%%%%%%%%%%%
\section{Model}
Consider $N$ hard spheres of diameter $\sigma=1$ in a planar 
slit with the surface area $S$. The slit itself is a part of 
a hosting system, whose coupling to the guest species changes
the slit geometry. Therefore, the slit width $h$ is not fixed, it 
fluctuates according to the fluid density (see below). 
For any $h$ the total Hamiltonian is   
\begin{equation}
H=H_{ff}+H_{fw},
\end{equation}
where $H_{ff}$ is the hard sphere Hamiltonian, $H_{fw}$ is the slit
potential  
\begin{equation}
H_{fw}=A\sum_{i=1}^N 
\left[
\frac{1}{z_i^k}+\frac{1}{(h-z_i)^k}
\right];
\qquad
1/2 \le z_i \le h-1/2.
\end{equation}
We do not take into account a short-ranged fluid-wall attraction,
responsible for the surface adsorption or layering effects. The 
inverse-power shape for $H_{fw}$ is chosen as a generic form of a soft 
repulsion, with $k$ controlling the softness. For technical purposes we are 
working with $k=3$. Moreover, our results are qualitatively insensitive to a 
particular choice of $k$. 

It should be noted that we do not discuss an adsorption mechanism 
or the equilibrium between the pore and the bulk fluids. In our case
$N$ is fixed and we focus on the pressure variation due to the changes
in the slit geometry.

%%%%%%%%%%%%%%%%%%%%%%%%%%%%%%
\section{Perturbation theory}
%%%%%%%%%%%%%%%%%%%%%%%%%%%%%%
In practice the evolution of the matrix morphology is much slower than the
fluid equilibration process. Then for a given width $h$ we can calculate
the fluid thermodynamics conditional to $h$. 
%%%%%%%%%%%%%%%%%%%%%%%%%%%%%%% 
\subsection{Conditional equation of state and insertion isotherm}
%%%%%%%%%%%%%%%%%%%%%%%%%%%%%%%%%
At any pore width $h$ the free energy can be represented as
\begin{equation}
\beta F(h)=\beta F_0(h)-\ln \langle e^{-\beta H_{fw}}\rangle_0
\end{equation}
where $F_0(h)$ is the free energy of a reference system, and 
$\langle...\rangle_0$ is the average over the reference state.  
Taking a spatially confined hard sphere system (the one 
with $A=0$) as a reference, we consider a first order 
perturbation\cite{Schoen,Kuz,pore} for the conditional free energy
\begin{equation}
\beta F(h)=\beta F_0(h) +\beta S \rho (h) \int \prod_{i} d{z}_i H_{fw} 
\end{equation}
where $\beta=1/(kT)$ and $\rho(h)$ is the pore density in the "slab"
approximation 
\begin{equation}
\label{rh}
\rho(h)=\frac{N}{S(h-1)},
\end{equation}
ignoring a non-monotonic behavior of the  
density profile with increasing pore density. This is reasonable for wide 
pores and low fluid densities. 
The reference part is estimated in the excluded volume approximation, while
the perturbation contribution is $N\Psi(h)$, such that the total conditional
free energy is
\begin{equation}
\label{Fh}
\beta F (h)=-N\ln{\left[
\frac{1-b\rho(h)}{\Lambda_T^3\rho(h)}
\right] 
} -N +N\beta\Psi(h)
\end{equation}
where $b=2\pi\sigma^3/3$ is the excluded volume factor, $\Lambda_T$ is
the thermal de Broglie length and
\begin{equation}
\label{Psi}
\Psi(h)=16A\frac{h}{(1-2h)^2}
\end{equation}
Tangential  $P_t(h)$ and normal $P_n(h)$ pressures can be found
as
\begin{equation}
P_t(h)=-\frac{1}{h-1}\left(\frac{\partial F(h)}{\partial 
S}\right)_{\beta, N}; \qquad
P_n(h)=-\frac{1}{S}\left(\frac{\partial F(h)}{\partial h}
\right)_{\beta, N}.
\end{equation}
This leads to
\begin{equation}
\beta P_t(h)=\frac{\rho(h)}{1-b\rho(h)}; \qquad
\beta P_n(h)=\frac{\rho(h)}{1-b\rho(h)}+
16 A^*\frac{N}{S}\frac{2h+1}{(2h-1)^3}.
\end{equation}
where $A^*=\beta A$.
Equation (\ref{rh}) allows us to eliminate the surface density $N/S$ in the 
favor of $\rho(h)$ and we obtain the following equation of state
\begin{equation}
\label{excluded}
\beta P_n(h)=
\frac{\rho(h)}{1-b\rho(h)}+
16 A^*\rho(h)\frac{(2h+1)(h-1)}{(2h-1)^3}
\end{equation} 
Therefore, for a fixed $h$, our result is clear and simple. The tangential 
pressure has the bulk form, with the bulk density being replaced by the
pore density. As a consequence of the first-order perturbative approach, 
$P_t(h)$ does not depend on the slit-fluid interaction. As we will see 
later, the simulation results demonstrate that this dependence is indeed 
quite weak. If necessary this effect can be reproduced theoretically taking
into account the second-order perturbation term.
 The normal pressure increases due to the repulsion $A^*$
and decreases with increasing pore width $h$, such that $P_n(h)=P_t(h)$
as $h\to \infty$.  

Recall that we are dealing with a fixed $N$. The insertion process
can be considered in the same framework. Just instead of the pressure 
components one would calculate the system response to increasing $N$--  
the insertion isotherm \begin{equation}
\beta \mu(h)=\left(\frac{\partial \beta F(h)}{\partial N} 
\right)_{\beta,S,h}=
\ln{\left[
\frac{\Lambda_T^3\rho(h)}{1-b\rho(h)}
\right] 
}+\frac{b\rho(h)}{1-b\rho(h)} + \beta\Psi(h)
\end{equation}
which relates the fluid density and the chemical potential $\mu(h)$. 
%%%%%%%%%%%%%%%%%%%%%%%%
\subsection{Dilatation effect}
%%%%%%%%%%%%%%%%%%%%%%%%
As is discussed above, the pore dilatation can be taken into account,
assuming that the width $h$ is density-dependent. Real insertion materials
are usually rather complex (multicomponent and heterogeneous). For that 
reason one usually deals with a distribution of pore sizes or with an
average size. In this context we assume that $h$ is known only statistically
and the probability distribution $f(h|\rho)$ is 
conditional\cite{Infoacta,SUSCINF} to the guest density $\rho$, which should 
be selfconsistently found from \begin{equation} 
\label{rp}
\rho=\int dh f(h|\rho) \rho(h). 
\end{equation}
Then, taking our results for $P_n(h)$ and $P_t(h)$, we can focus on the 
equation of state averaged over the width fluctuations.  
\begin{equation}
\label{p}
P_i=\int dh f(h|\rho) P_i(h); \quad i=n,t.
\end{equation}

This, however, requires a knowledge on the distribution $f(h|\rho)$.
Even without resorting to a concrete form for $f(h|\rho)$, it is
clear that the matrix reaction can be manifested as a change in
the distribution width or/and the mean value.  
 One of the simplest forms
reflecting at least one of these features is a $\delta$-like 
distribution, ignoring a non-zero width. 
\begin{equation}
f(h|\rho)=\delta[h-h(\rho)]
\end{equation}
where $\delta(x)$ is the Dirac $\delta$-function and the mean pore width
$h(\rho)$ is density-dependent. 
\begin{equation}
\label{hr} 
h(\rho)=h_0(1+\tanh[\Delta (\rho-\rho_0)]+\tanh[\Delta \rho_0])
\end{equation}
This form mimics a non-Vegard behavior,
typical for layered intercalation compounds\cite{SOSI}. At low densities 
($\rho << \rho_0$) the dilatation is weak. The most intensive 
response is at $\rho \approx \rho_0$, and then the pore reaches a 
saturation, corresponding to its mechanical stability limit. Here  $\Delta$ 
is the matrix response constant or dilatation rate, controlling the 
slope near $\rho \approx \rho_0$. 

From eq.~(\ref{rp}) the average density is found 
to be 
\begin{equation}
\rho=\frac{N}{S}\frac{1}{h(\rho)-1}
\end{equation}
Changing the surface density $N/S$ we vary the average pore density
$\rho$. This allows us to eliminate $N/S$ in the favor of $\rho$ in all 
thermodynamic functions. Combining eqs~(\ref{excluded}) and (\ref{p}) we 
obtain \begin{equation}
\beta P_n=\frac{\rho}{1-b\rho}
+
16 A^* \rho\frac{(2h(\rho)+1)(h(\rho)-1)}{(2h(\rho)-1)^3}; \qquad
\end{equation}
It is convenient to introduce the compressibility function
\begin{equation}
\chi_n=\frac{1}{h(\rho)-1}\frac{\partial h(\rho)}{\partial P_n}=
\frac{1}{h(\rho)-1}\frac{\partial h(\rho)}{\partial \rho}
\frac{\partial \rho}{\partial P_n}
\end{equation}
which vanishes in a non-swelling limit ($h(\rho)=h_0$). It is clearly seen
that $\chi_n$ could be singular either if the swelling is discontinuous,
or if $P_n$ exhibits an inflection point. In what follows we focus on the
second option.

In the case of a wide and weakly reacting pore we expand in terms
of $1/h_0$ and $\Delta$, obtaining a generic van der Waals form  
\begin{equation}
\beta P_n=\frac{\rho}{1-b\rho}+\frac{4A^*}{h_0}\rho 
-\frac{4A^*\Delta}{h_0}\rho^2
\end{equation}
Therefore, we have an interplay of several effects -- the packing (first 
term), the fluid-matrix interaction (linear in density), and the matrix 
reaction (quadratic term).  It is seen that 
a coupling of the fluid-slit repulsion ($A$), spatial 
 confinement ($h_0$) and the matrix response ($\Delta$) acts as an effective 
 infinite-range fluid-fluid attraction (but just in one direction). 
Introducing a dimensionless temperature $T^*=1/(4A^*)$, and solving
\begin{equation}
\frac{\partial P_n}{\partial \rho}=0; \qquad
\frac{\partial^2 P_n}{\partial \rho^2}=0
\end{equation} 
we find the "critical"
parameters
\begin{equation}
\rho_c=\frac{1}{3}\frac{\Delta+b}{\Delta b}; \qquad
T^*_c=\frac{1}{27}\frac{(2 \Delta-b)^3}{ b h_0 \Delta^2}
\end{equation}
at which the normal compressibility $\chi_n$ diverges.
It is seen that a physically meaningful (with $T^*_c>0$) pseudo-criticality 
appears only at $\Delta>b/2$. In other words, the matrix reaction $\Delta$
should dominate the packing effects.  
In addition,  $T^*_c$ decreases  with increasing pore width $h_0$. 

It is well known that for bulk systems the vdW loop appearing at $T^*<T^*_c$
is unphysical and one usually invokes the Maxwell construction, determining
the liquid-vapor coexistence. As we will see below, in our case the loop
is a physically justified effect. It appears as a competition between
the packing, which tends to increase the pressure with increasing $\rho$ and
the slit dilatation,  decreasing $P_n$ with increasing $\rho$. As a
result the states with negative linear compressibility are stabilized. 
Note that this behavior is not sensitive to the particular form (\ref{hr})
of $h(\rho)$. Moreover, the dilatation rate $\Delta$ is essential, while
the dilatation magnitude $h(\rho)-h_0$ plays only a marginal role.  
 As it  should be, this effect disappears in the bulk limit $h_0 \to \infty$ 
 or in the case of insensitive matrices $\Delta \to 0$. 

In order to be more accurate in comparison to the MC data the reference
part is replaced by the Carnahan-Starling form 
\begin{equation}
\label{CS}
\beta P_n=\rho\frac{1+\eta+\eta^2-\eta^3}
{[1-\eta]^3}+
16 A^* \rho\frac{(2h(\rho)+1)(h(\rho)-1)}{(2h(\rho)-1)^3}; \qquad
\eta=\pi \rho/6.
\end{equation}
This allows us to avoid the unphysical behavior at high densities.

%%%%%%%%%%%%%%%%%%%%%%%%%%%
\section{Simulation}
%%%%%%%%%%%%%%%%%%%%%%%%%%
In order to verify the existence of the loop predicted by the perturbation 
theory as well as to get more insight into the phenomenon the  
MC  simulation of the model has been carried out. We applied the canonical 
NVT MC simulations of a confined hard sphere fluid to calculate the density 
profiles and pressure components. The simulation cell was parallelepiped in shape, 
with parallel walls at surface separation $h$, and surface area $S=L_x 
\times L_y$. The periodic boundary conditions were applied to the $X$ and 
$Y$ directions of the simulation box; the box length in the $Z$ direction 
is fixed by the pore width. For a given pore width the adsorbed fluid 
density is chosen according to the Eq.(\ref{rh}). There was constant number 
of particles, $N=1000$, and a desired fluid density has been get by 
adjusting the value of area, $S$. We repeated our simulation runs with 
bigger number of particles, $N =1500$ and $3000$, but no 
significant differences were found. The density profiles, $\rho (z)$, were 
calculated in the usual way by counting the number of particles, $N_i$ in 
the slabs of thickness $d_z=0.05$ parallel to the plane $XY$ by using $\rho 
(z)= N_i/v$, where $v$ is the slab volume $v=d_z \times L_x \times L_y$. The 
definition of Irving and Kirkwood \cite{irving} has been used to calculate 
the components of the pressure. The normal component of the pressure for the 
fluid-fluid interaction is 
\begin{equation} 
\label{Pn}   \beta P_n (z) = \rho(z)  - 
\frac{\beta}{S}\left \langle  \sum_i \sum_{j>i}\frac{dH_{ff}}{dr} 
\frac{z_{ij}^2}{ |r_{ij}|} \Theta (\frac{z-z_i}{z_{ij}}) \Theta 
(\frac{z_j-z}{z_{ij}} ) \right \rangle 
\end{equation}
where  $\rho(z)$ is the total density profile and $\langle...\rangle$ 
denotes the average over the MC configurations. The tangential component, 
$P_t(z)$, has been calculated by substitution of $z_{ij}^2$ by   
$0.5(x_{ij}^2+y_{ij}^2)$ in equation (\ref{Pn}). For hard spheres the 
derivative of the potential is
\begin{equation} 
\label{Delta} 
\frac{d\beta H_{ff}}{dr}=- \delta(r-1) 
\end{equation}

The Dirac $\delta$ function in our simulation 
was approximated as
\begin{equation}
\label{Lam}
\delta (r-1)=\frac{\Theta (r-1) 
-\Theta(r-1-\Lambda)}{\Lambda},
\end{equation}
as $\Lambda \to 0$. In Eq. (\ref{Lam}) $\Lambda$ 
is used to define the region where two particles collides. A collision 
between particles occurs if $ 1 < r < 1 + \Lambda $. Therefore, 
for the particular case of hard spheres the averaged normal component of 
the pressure tensor for fluid-fluid interactions can be reduced 
to
\begin{equation}
\label{Pnav}
\langle  \beta P_n \rangle = \langle \rho \rangle  + \frac{1}{V}  
\left \langle \sum_i 
\sum_{j>i} \frac{1}{\Lambda} \frac{z_{ij}^2}{ |r_{ij}|} \Theta 
(\frac{z-z_i}{z_{ij}}) \Theta (\frac{z_j-z}{z_{ij}}) 
\right \rangle 
\end{equation}
The wall-fluid interaction is a function of 
$z$ and affects only $P_n$ while the tangential component remains the same. 
The normal pressure in this case is
\begin{equation}
\label{PnW1}
\beta P_n^W(z) = \beta P_n^{W1}(z)  + \beta P_n^{W2}(z)
\end{equation}
where $\beta P_n^{W1}(z)$ and $\beta P_n^{W2}(z)$   are the fluid-wall contributions 
from surfaces located at $z_{W1} = 0.5$   and $z_{W2} = h - 0.5$, 
respectively. These contributions are defined as
\begin{equation}
\label{W1}
\beta P_n^{W1}(z) = - \frac{\beta}{S} \left \langle \sum_i \frac{dH^i_{fw}(z)}{dz}|z_i-z_{W1}|
\Theta (z_i-z) \Theta (z-z_{W1}) \right \rangle 
\end{equation}
\begin{equation}
\label{W2}
\beta P_n^{W2} (z) = - \frac{\beta}{S} \left \langle 
\sum_i \frac{dH^i_{fw}(z)}{dz} |z_{W2}-z_i| \Theta (z_{W2}-z) \Theta 
(z-z_i) \right \rangle
\end{equation}

These definitions  of the pressure tensor treat  
the wall-particle interaction as a contribution to the intermolecular forces 
in a system consisting of the fluid and solid, rather than as an external 
field acting on the fluid. Thus the normal component of the pressure tensor 
must be independent of $z$ as a condition of mechanical equilibrium and for 
a system containing bulk fluid must be equal to the pressure at the bulk 
density. Tangential components  should approach the bulk pressure for 
sufficiently large $h$.

In this work we have taken parameter $\Lambda $  
equal to $0.001$, $0.002$, and $0.003$ to make the extrapolation. The 
average normal component,includes the fluid-fluid and fluid-wall 
contributions given by equations (\ref{Pnav}) and (\ref{PnW1}). Each 
simulation runs $5\times 10^5$ MC cycles, with the first half for the 
system to reach equilibrium whereas the second half for evaluating the 
ensemble averages.

%%%%%%%%%%%%%%%%%%%
\section{Results}
In agreement with our theoretical prediction, the simulation results confirm 
that the negative compressibility states (the loop) appear only if the 
slit reaction $\Delta$ reaches some threshold value $\Delta^*$ which 
involves a combination of $h_0$, $\rho_0$ and $A$. The variation of
the average density $\rho$ due to the dilatation should dominate its
variation, induced by the changes in the surface density $N/S$.  

Pressure as a function of the average pore density is plotted in Figure~1.
It is seen that the normal component $P_n$ develops the loop with
increasing pore-fluid repulsion $A$. The simulation results confirm
that this feature is not an artefact of our theoretical approximations.
As expected\cite{pore}, the perturbation theory systematically 
underestimates the magnitude of the fluid-wall repulsion, $A$, 
(overestimates the temperature $T^*$) at which this effect takes place. 
The tangential component $P_t$ is  much
less sensitive to the slit dilatation. This is coherent with our theoretical
estimation. Interestingly, that $P_t$ can be reasonably fitted  by the
expression for $P_n$ taken at much lower $A$ (see the insets).  
Therefore, we expect that at sufficiently strong repulsion
(low temperatures) $P_t$ could also be non-monotonic. This would correspond
to a fluid-solid transformation, which is not considered here. 
The loop becomes more pronounced with increasing dilatation
rate $\Delta$. Since only one pressure component exhibits this behavior, 
there is 
no rational basis to suspect a vdW instability (of 
liquid-vapor type). Moreover, we are dealing with rather wide pores
($h_0=10$) in order to avoid the narrow channel effects\cite{Mukamel}. 
This makes us to search for an alternative explanation.  

With this purpose we have analyzed the fluid structure at different
densities. The density profiles are presented 
in Figure~2. It is obvious that the system does not exhibit strong density 
oscillations. It is worth noting, that our perturbation theory results are 
obtained in the
'slab' approximation, eq.(\ref{rh}), which does not take them into account. 
On the other hand, analysis of the density profiles indicates  
a sudden change of the packing regime in 
the negative compressibility region (in between the points $A$ and $B$, 
marked in the Figure~1~(b)). The fluid film becomes more dilute with 
decreasing (e. g. due to a lateral stretching) density 
up to the point $B$ with $\rho=0.32$. This process goes uniformly in the 
middle of the film and at the periphery. Passing from $\rho=0.32$ to 
$\rho=0.29$ does not change the middle density, while the rarefaction 
takes place only near the slit walls. Upon reaching $\rho=0.25$ the film 
suddenly densifies in the middle. This corresponds to the inflection point 
in Figure~1~(b). Up to the point $A$ with $\rho=0.2$ the film dilutes only 
at its periphery. Then we return to the usual uniform dilution with further 
decrease in the density. Therefore, there is a clear 
correlation between the negative compressibility states and the packing 
mechanism. 

In order to study the redistribution of the fluid density we have calculated 
the average density $\Gamma(z)$ in slabs of different thickness $z$
\begin{equation}
\Gamma(z)= \int_0^z \rho (t) dt
\end{equation}
The obtained results  are presented in Figure~3. As seen,  $\Gamma(z)$ 
exhibits the same loop as the normal pressure does. Going towards the 
middle of the pore makes this effect more pronounced.    
Such similarity permits 
us to conclude that the fluid-wall soft repulsion (coupled to the 
dilatation $\Delta$) is a main reason of the fluid reordering inside the 
pore, and as a consequence the states with negative compressibility can 
occur. 
This is clearly seen from eq.~(\ref{CS}), that 
gives a monotonic isotherm as $A \to 0$ or  $\Delta \to 0$.

It is interesting to mention that a similar trend has been described 
recently in the studies of mineral clays swelling   
\cite{ODRI,SMITH}.
Namely, Smith et al. \cite{SMITH} reported the simulation results of 
hydrated Na-smectites with variable layer charge. They found that the 
tendency to swell increases with increasing layer charge (increasing 
repulsion), which is consistent with our  conclusions.

%%%%%%%%%%%%%%%%%%%%%%%%%
%%%%%%%%%%%%%%%%%%%%%%%%%
\section{Conclusion}
%%%%%%%%%%%%%%%%%%%%%%%%
We have found  that a coupling of the fluid-slit 
repulsion, spatial confinement and the slit dilatation acts as an 
anisotropic fluid-fluid attraction, inducing negative linear compressibility 
states. These states are shown to be related to an abrupt change in
the packing regime, including a local densification in the middle
with decreasing surface density.  This resembles the stretch-densification
effects\cite{Baughman1,Baughman2} in negative compressibility materials.
In the context of our model the mechanism is the following. Stretching
in the lateral direction decreases the surface $N/S$ and the pore $\rho$
densities. This leads to a transversal shrink $h(\rho)$ at the
rate $\Delta$. This stabilizes the negative transversal compressibility
when the rate passes some threshould value. This mechanism is quite 
different from that explored in the low-dimensional 
systems\cite{Two1,Two2,Two3,Mukamel}, where the loops appeared essentially 
due to the small-size effects restricting the phase space accessibility when
the box length became comparable with the particle size. As a result,
the vdW feature is found\cite{Two3} to be very sensitive to the box geometry 
(rectangular or spherical). In our
case we deal with a three-dimensional system where the particles can 
exchange their positions almost freely ($10\le h(\rho)/\sigma \le30$), 
although the phase space is somewhat restricted by the slit-fluid repulsion.
Nevertheless, our model shares some of the low-dimensional 
features\cite{Two1,Two2,Two3,Mukamel} and we recover the usual monotonic
behavior in the bulk limit $h_0 \to \infty$. 

Since the tangential pressure component does not exhibit this effect, the
system does not undergo a phase transition (at least in its traditional 
sense). This is confirmed by the absence of strong density fluctuations (in 
both directions). On the other hand, the loop appears in the direction, along
which our system is finite (finite $h$) and, therefore, the negative 
compressibility can be considered as a finite-size effect\cite{Gulminelli},
vanishing in the bulk limit.  
Nevertheless, an inclusion of an attractive 
fluid-fluid interaction would result\cite{Schoen,Kuz} in the pore 
condensation. In this respect it would be quite interesting to analyze how 
the above stretch-densification coexists with the true critical behavior. 
The role of more isotropic geometries (e.g. spherical pores) as well as
that of a non-zero distribution width ($f(h|\rho)$) and the potential 
softness $k$ could also be discussed. 
Another interesting point is to describe the sorption behavior, that is,
changing $\rho$ by varying $N$ at fixed $S$. In this way  we can 
mimic a "dosen" adsorption\cite{Amarasekera} or the equilibrium with an 
infinite bulk reservoir. 
We plan to analyze these issues in a future work.  
 
%%%%%%%%%%%%%%%%%%%%%%%%%%%

%%%%%%%%%%%%%%%%%%%%%%%%%%%%%%%%
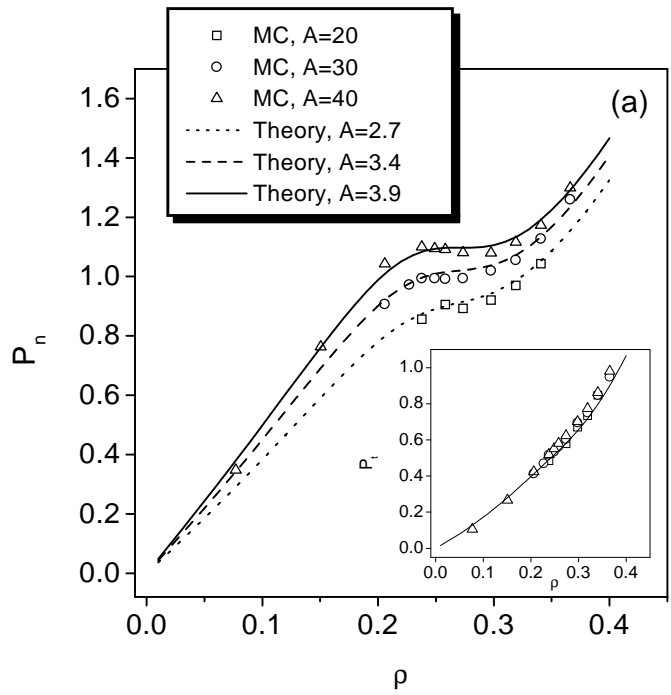
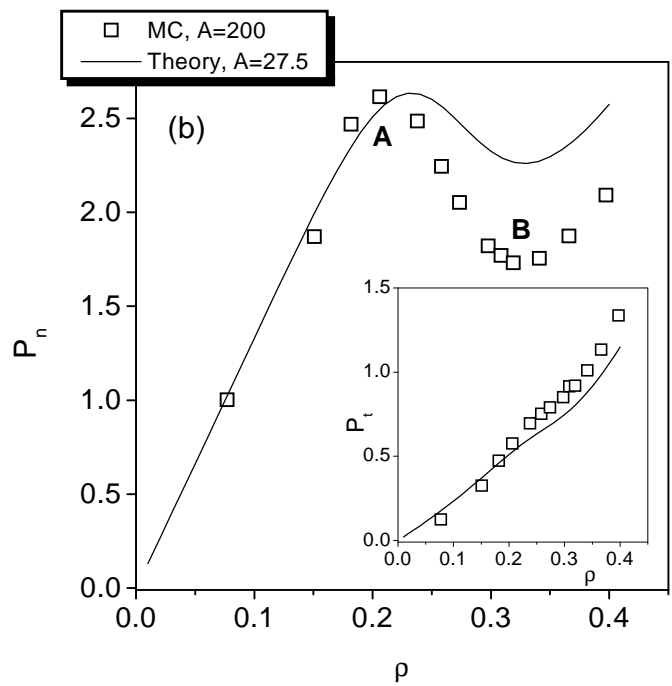
\begin{figure}
\caption{Normal and tangential pressures (the insets) in the case of
a weakly (a) and strongly (b) repulsive slit. The other parameters are
$h_0=10$, $\Delta=15$, $\rho_0=0.3$} 
\end{figure}
%%%%%%%%%%%%%%%
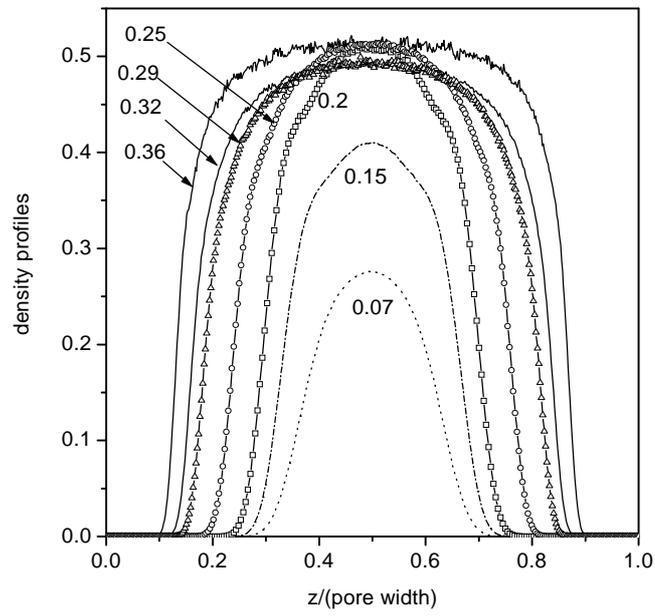
\begin{figure}
\caption{MC  density profiles  at different densities (indicated). Other
parameters as in the previous figure. }
\end{figure}
%%%%%%%%%%%%%%%%%%%%
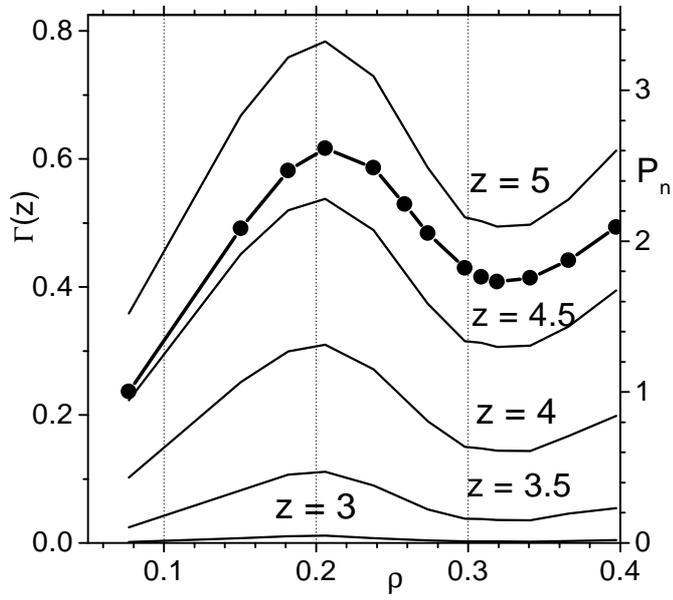
\begin{figure}
\caption{Average fluid density in slabs of different thickness $z$ (lines) 
and the normal pressure component (line with symbols).} 
\end{figure}
%%%%%%%%%%%%%%%%%%%%%%%%%%%
%\begin{figure}
%\caption{}
%\end{figure}

%%%%%%%%%%%%%%%%%%%%%%%
 \end{document}